\documentclass[fleqn,usenatbib,useAMS]{mnras}

\usepackage{graphicx}	
\usepackage{amsmath}	
\usepackage{amssymb}	
\usepackage{multicol}        
\usepackage{bm}		
\usepackage{pdflscape}	

\defcitealias{Facchini:2017of}{F18}

\usepackage[T1]{fontenc}
\usepackage{ae,aecompl}

\usepackage{newtxtext,newtxmath}

\title[Rocking shadows]{Rocking shadows in broken circumbinary discs}

\author[R. Nealon et al.]{\parbox{\textwidth}{
Rebecca Nealon$^{1}$\thanks{corresponding author: rebecca.nealon@leicester.ac.uk},
Daniel~J. Price$^{2}$ and
Christophe Pinte$^{2}$}\vspace{0.15cm}\\
$^{1}$Department of Physics and Astronomy, University of Leicester, Leicester, LE1 7RH, United Kingdom,\\
$^{2}$School of Physics and Astronomy, Monash University, Clayton Vic 3800, Australia}

\date{Last updated \today; in original form \today}

\pubyear{2020}

\begin{document}
\label{firstpage}
\pagerange{\pageref{firstpage}--\pageref{lastpage}}
\maketitle

\begin{abstract}
We use three dimensional simulations with coupled hydrodynamics and Monte Carlo radiative transfer to show that shadows cast by the inner disc in broken circumbinary discs move within a confined range of position angles on the outer disc. Over time, shadows appear to rock back and forth in azimuth as the inner disc precesses. The effect occurs because the inner disc precesses around a vector that is not the angular momentum vector of the outer disc. We relate our findings to recent observations of shadows in discs.
\end{abstract}




\section{Introduction}

Narrow lane shadows have been observed in scattered light observations of a number of protoplanetary discs \citep{Marino:2015rh,Walsh:2017ic,Benisty:2018ve,vanderPlas:2019gy}. These indicate a large ($\sim 70-90 \degr$ relative misalignment between the inner and outer disc \citep{Marino:2015rh,Walsh:2017ic,Pinilla:2018gb}. 

These disc structures are likely to be caused by a massive companion on an orbit that is misaligned to the outer disc. Numerical simulations have shown that the presence of a misaligned companion can `break' an accretion disc into an inner and outer disc, which then differentially precesses \citep{nixon_2013,Facchini:2017of,Nealon:2018ic,Zhu:2018vf}. This can lead to relative misalignments greater than $90 \degr$ (\citealt{Facchini:2017of}; hereafter \citetalias{Facchini:2017of}). To date, for protoplanetary discs these models require a massive companion (an equal mass binary in \citetalias{Facchini:2017of} or a $\gtrsim$10 M$_{\rm J}$ planet in \citealt{Zhu:2018vf}). With the exception of HD~142527 \citep{Biller:2012ct,Lacour:2016ij}, these companions are as yet undetected.


Ideally, we would like to infer inner disc properties from the observed shadows. \citetalias{Facchini:2017of} investigated the observational signatures by post-processing hydrodynamic simulations, finding that shadows can affect the temperature profile in the outer disc (see also~\citealt{Casassus:2019pt}). \citet{Min:2017oc} showed that the illumination pattern can tightly constrain the relative geometry between the inner and outer disc. \citet{Zhu:2018vf} extended this by considering the evolution of the shadow as the inner disc precessed by $60 \degr$. They showed that the shadow does not move at a constant rate around the disc --- instead, the shadow can rotate faster or slower than the inner disc, depending on the phase of the precession. But these simulations did not account for temperature changes in the disc caused by the shadow or study the long-term evolution of the shadows.

Here we consider the evolution of shadows on longer timescales. We simulate a broken circumbinary disc using parameters based on \citetalias{Facchini:2017of} but with a less massive companion. Unlike \citetalias{Facchini:2017of} we account for temperature changes caused by the shadow. We show that the shadows cast by the misaligned inner disc rock rather than roll around the whole outer disc in synthetic scattered light images. Our shadows are reminiscent of those observed in the disc around 2MASS J16042165-2130284 (hereafter J1604), suggesting the presence of a precessing inner misaligned disc.

\begin{figure*}
    \centering
    \includegraphics[width=\textwidth]{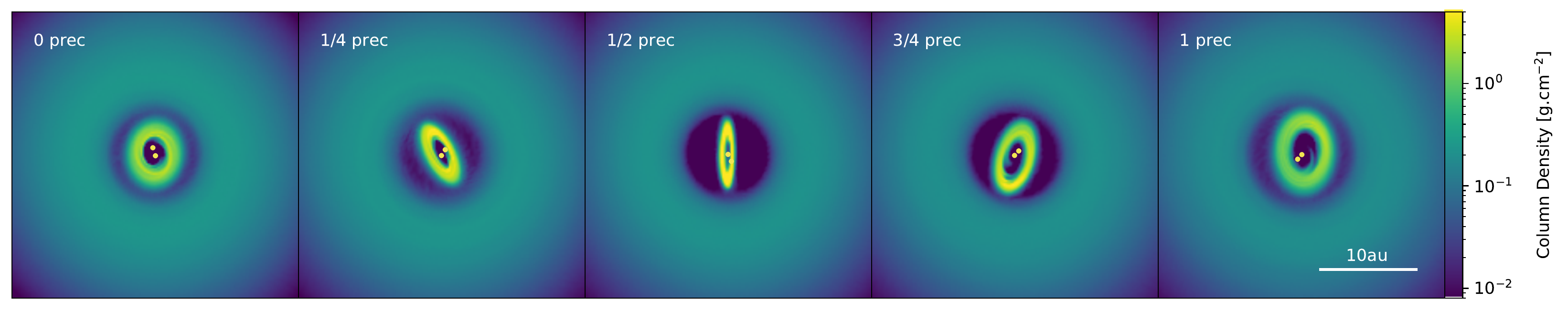}
    \includegraphics[width=\textwidth]{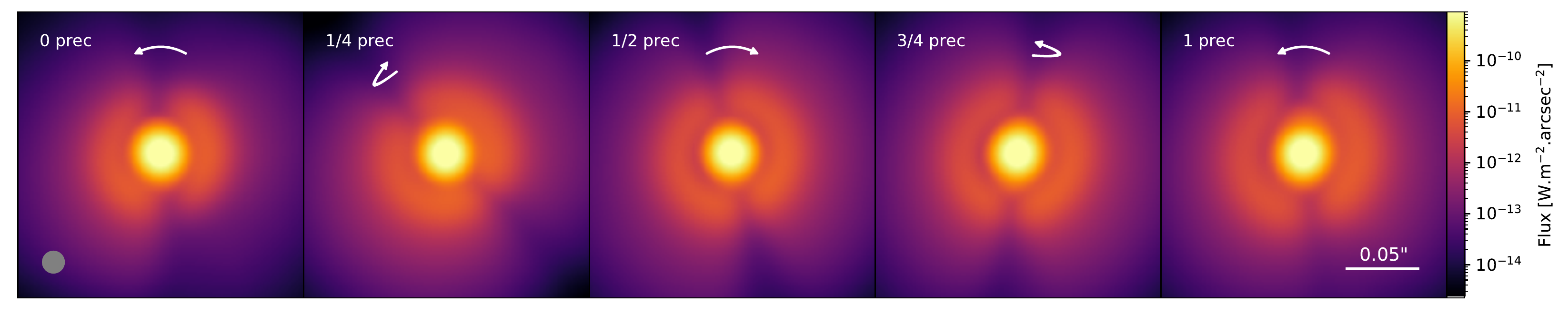}
    \includegraphics[width=\textwidth]{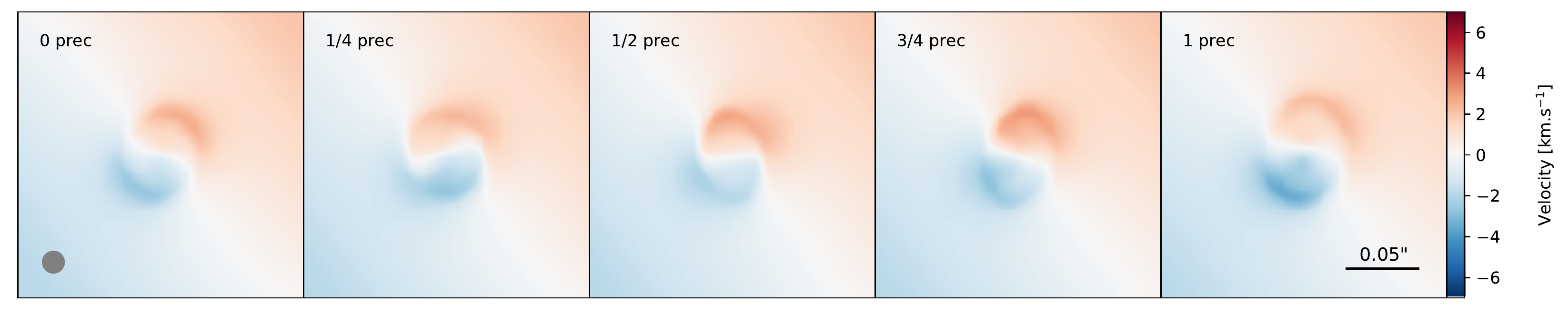}
    \caption{Rocking shadows in broken circumbinary discs over one precession period of the inner disc, assuming a binary orbital plane misaligned to outer disc by 60 degrees. Observer is face-on to the outer disc. Upper row: column density. Middle row: synthetic scattered light images. White arrows indicate instantaneous direction of motion of the shadow and the grey circle in first panel shows simulated beam width of 15mas. Lower row: CO (3-2) Moment 1 map. Times shown in panels correspond to dotted lines in Figures~\ref{fig:disc_with_time} and the panels in Figure~\ref{fig:temperature_profiles}. All panels are shown on the same spatial scale, indicated in the right-most panel of each row.}
    \label{fig:density_sl}
\end{figure*}

\section{Method}
\label{sec:methods}
We perform 3D simulations using the coupled version of \textsc{phantom+mcfost}. That is, we evolve the hydrodynamics using the \textsc{phantom} smoothed particle hydrodynamics (SPH) code \citep{Phantom}, with temperature during the simulation computed using the \textsc{mcfost} Monte Carlo radiative transfer code \citep{Pinte:2006nw,Pinte:2009ye}. This coupling allows us to take into account the effect of the shadows that are cast onto the disc due to any misaligned gas in the inner regions.

We model a circumbinary disc with a binary separation of 1~au using sink particles with an accretion radius of 0.45~au. The primary and secondary mass are set to $M_{\rm 1} = 1.0\,M_\odot$ and $M_{\rm 2} = 0.1\,M_\odot$ such that the mass ratio, $q=0.1$ is consistent with limits from typical observations \citep[e.g.][]{Benisty:2018ve}. The disc is initially modelled between $R_{\rm in}=1.7$~au and $R_{\rm out}=15$~au with a total mass of $1\times 10^{-4}\,M_\odot$ (where $R$ is the cylindrical radius). The surface density profile is described by a power-law with $R^{-1}$ and a taper at the inner edge. We set the initial aspect ratio to $H/R=0.05$ at $R_{\rm in}$, but the vertical disc structure then evolves self-consistently based on the temperature. Finally, viscosity in the disc is implemented following \citet{Flebbe:1994lr} with $\alpha=0.02$ and the artificial viscosity $\alpha_{\rm AV}$ varying between 0.1 and 1.0 based on the presence of shocks \citep[see][for further details]{Phantom}. With the exception of the binary and disc mass, these initial parameters are almost identical to \citetalias{Facchini:2017of}. We model the disc using $1\times 10^6$ particles, corresponding to more than one smoothing length per scale-height \citepalias[Figure~1,][]{Facchini:2017of}.

We assume radiative equilibrium. That is, we neglect time dependence in the radiative transfer and assume that absorbed energy is instantly re-emitted. To illuminate the disc and set the temperature, we update the temperature from \textsc{mcfost} every 0.707 $\times$ the binary orbit (so that the stars are not in the same position each time). We use $10^8$ photon packets on a Voronoi mesh built around the SPH particles. We compute the luminosity of the stars assuming a 3 Myr isochrone from \citet{Siess:2000vd}, with corresponding temperatures of $T_1=4262$K and $T_2=2948$K and using their updated masses from the simulation. As the observational signatures we are looking for are predominantly in scattered light, we do not include dust in the {\sc phantom} simulations. Instead, we simply assume that the dust and gas are in thermal equilibrium and a constant gas/dust ratio of 100. The dust grains are spherical and homogeneous, with an opacity that is independent of temperature. The dust grains are distributed across 100 grain sizes between 0.03 and 1000$\mu$m with a power-law exponent of -3.5 \citep{Mathis:1977bw}. To calculate the moment maps we assume a uniform CO-to-H2 ratio of $1 \times 10^{-6}$.

We start with the binary and disc angular momentum vectors aligned and allow the disc to evolve until the temperature structure (indicated by $H/R$) reaches a steady state. For our parameters, this corresponds to $\sim280$ binary orbits. The gas disc was then rotated through a misalignment of $60\degr$ and the evolution continued for a subsequent 1500 binary orbits. We then use \textsc{mcfost} to calculate the synthetic scattered light images at $1.6\mu$m from various snapshots of the simulation in post-processing (assuming the stellar parameters above). The scattered light images assume a distance of 150pc and that the plane of the outer disc is aligned with the plane of the sky. We convolve the images with a 2D Gaussian using a small beam of 15mas as our system is quite compact (but this choice does not affect the evolution of the shadows).

To measure the position of the shadows we azimuthally binned the flux profile near the inner edge of the outer disc between 8 and 10~au for each scattered light image (e.g. in the middle panel of Figure~\ref{fig:density_sl}). We then normalise this azimuthal profile by its maximum and identify the location of the shadows by measuring the positions of the local minima.

\section{Results}
\begin{figure}
    \centering
    \includegraphics[width=\columnwidth]{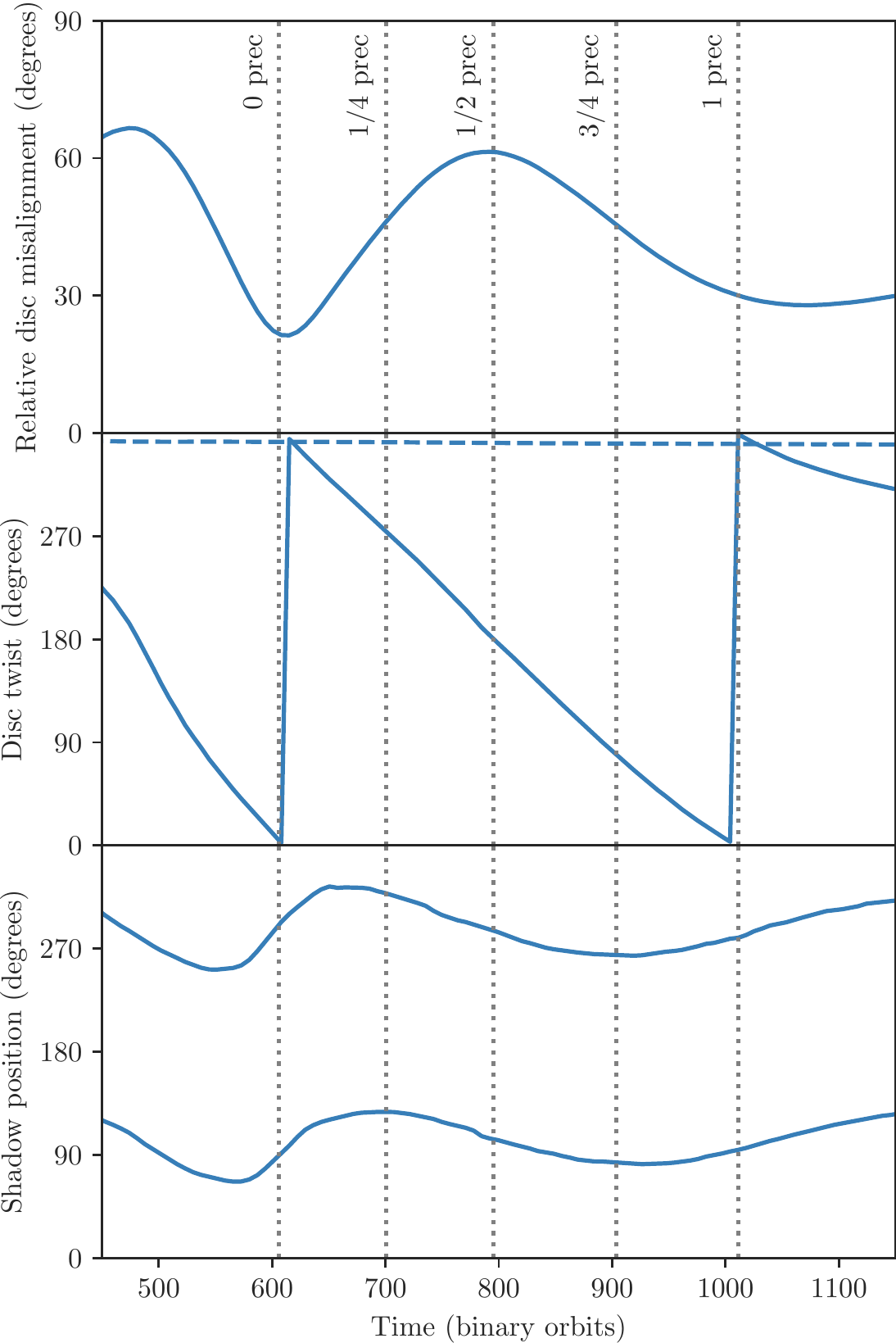}
    \caption{Evolution of the disc misalignment. Upper panel: Relative misalignment between inner and outer disc, measured as the angle between the weighted average angular momenta vectors of the discs. Middle panel: Twist angle traced out around the total angular momentum vector by the angular momentum vector of the inner disc measured at 3.5~au (solid line) and the outer disc at 12~au (dashed line). Lower panel: azimuthal location of the shadows, measured from the scattered light images. Vertical lines indicate 1/4 of a precession cycle and correspond to time of each panel in Figure~\ref{fig:density_sl}.}
    \label{fig:disc_with_time}
\end{figure}

The upper row of Figure~\ref{fig:density_sl} shows the time evolution of column density over one precession period. Following \citetalias{Facchini:2017of}, we find that the misaligned binary causes the disc to break into an inner and outer disc that differentially precess. Our disc evolution is similar to \citetalias{Facchini:2017of}, but takes longer to break due to the lower mass ratio (about 250 orbits once the disc is misaligned to the binary). The upper and middle panels of Figure~\ref{fig:disc_with_time} shows the evolution of the disc over 400 to 1200 binary orbits. We find a smaller maximum relative misalignment as in previous studies (compare upper panel of Figure~\ref{fig:disc_with_time} to Figure~7 of \citetalias{Facchini:2017of}).

The middle panel of Figure~\ref{fig:disc_with_time} shows the twist of the inner disc (measured at 3.5~au) and the outer disc (measured at 12~au), displaying constant retrograde precession (retrograde being the expected direction of precession around a binary, see e.g.~\citealt{Bate:2000fk}). Following \citetalias{Facchini:2017of}, we compare the precession timescale and break radius of our simulation with theoretical expectations. We measure the initial break radius at $3.5$~au, between the two limiting estimates of $1.1$~au and $4.4$~au \citep{nixon_2013}. Additionally, the innermost radius is measured at $R_{\rm in} = 1.5$~au and the average misalignment of the inner disc to be $\beta = 50\degr$ during the first precession of the inner disc. From Equation 4 of \citetalias{Facchini:2017of}, the predicted precession timescale of $434$ binary orbits is consistent with the $\sim 405$ measured from Figure~\ref{fig:disc_with_time}.

The middle row of Figure~\ref{fig:density_sl} shows the corresponding scattered light images. As in \citetalias{Facchini:2017of} and \citet{Zhu:2018vf} we recover narrow lane shadows due to the strong relative misalignment of the inner disc. White arrows on each panel indicate the instantaneous direction that the shadow is moving. The inner disc shadows rock back and forth around the disc on the precession period, while remaining confined to a narrow range of position angles ($\lesssim 30\degr$ from vertical in Figure~\ref{fig:density_sl}). The lower panel of Figure~\ref{fig:disc_with_time} shows the azimuthal position of the shadow measured from the scattered light images. In this representation the `back and forth' rocking of the shadow is apparent and occurs with the same period as the precession of the inner disc.

The lower panel of Figure~\ref{fig:density_sl} shows the Moment 1 of the CO J=3-2 line for the corresponding snapshots. Interior to the inner edge of the outer disc we identify the twisted kinematics characteristic of misaligned inner discs \citep[e.g.][]{Marino:2015rh,Casassus:2015yu,Casassus:2018te}, also seen in J1604 \citep{Mayama:2018yd}.

\begin{figure}
    \centering
    \includegraphics[width=\columnwidth]{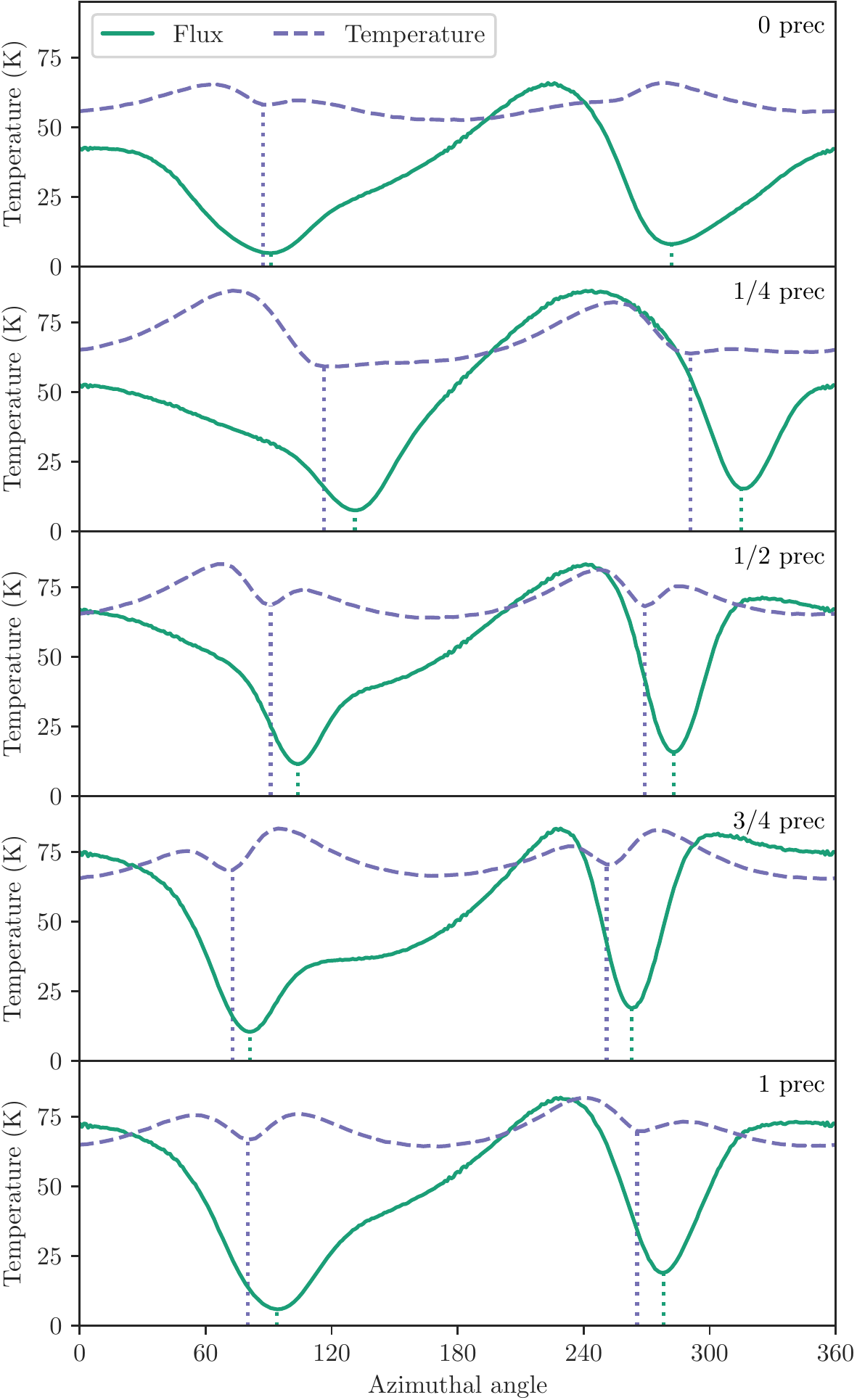}
    \caption{Azimuthal temperature (purple) and flux (green) profiles, scaled so that the maximum in flux coincides with the maximum temperature. Vertical dotted lines show the positions of the local minima (i.e. for the flux, the locations of the shadows). The flux profiles can be compared directly to figure~5 of \citet{Pinilla:2018gb}. Panels here correspond to time of each panel in Figure~\ref{fig:density_sl}.}
    \label{fig:temperature_profiles}
\end{figure}

Figure~\ref{fig:temperature_profiles} shows the azimuthal flux profile (also between 8 and 10~au) for the snapshots shown in Figure~\ref{fig:density_sl}.  Across a full precession of the inner disc the amplitudes, widths, location and separation between the shadows varies strongly. We note the similarity of the shadow properties (i.e. the dips) and asymmetry between the shadows in our flux profiles here compared to Figure~5 of \citet{Pinilla:2018gb}.

The azimuthal temperature profile is overplotted in Figure~\ref{fig:temperature_profiles}, scaled such that the maximums for both profiles coincide. Despite relatively large changes in the azimuthal flux profile, the temperature is only perturbed at the location of the narrow lane shadows. Figure~\ref{fig:temperature_profiles} also shows the location of the shadows and temperature minima. We find the temperature dips move in unision with the shadows cast by the misaligned inner disc. However, as the shadow rocks around the outer disc the corresponding dip in temperature may either lead or follow the shadow.

\section{Discussion}
The phenomenon of `rocking' shadows is a geometric effect that occurs because the inner disc precesses about a vector that is misaligned with respect to the angular momentum of the outer disc. To illustrate this, consider the extreme case where this reference vector is perpendicular to the angular momentum of the outer disc such that the inner disc is (on average) perpendicular with the outer disc. As the inner disc precesses, the longitude of the ascending and descending nodes oscillate forwards and backwards with a range that corresponds to the misalignment between the inner disc and the reference vector. As the shadow cast by the inner disc follows the line of nodes, this results in a rocking shadow.

Whether this effect occurs depends on the relative misalignment of the inner disc ($\beta_{\rm ID}$) and the outer disc ($\beta_{\rm OD}$) to the total angular momentum of the disc and binary. Rocking shadows occur when $\beta_{\rm ID} < \beta_{\rm OD}$ (i.e. the outer disc is more strongly misaligned). In our simulation shown in Figure~\ref{fig:density_sl}, the presence of a binary misaligned with the outer disc such that disc breaking can occur guarantees this geometry. Indeed, Figure~\ref{fig:disc_with_time} suggests that $\beta_{\rm OD} \sim 50\degr$ and $\beta_{\rm ID} \sim 20\degr$. Conversely, when $\beta_{\rm ID} > \beta_{\rm OD}$, the shadows cast will traverse the entire outer disc. The limiting case occurs when $\beta_{\rm ID} \sim \beta_{\rm OD}$, where shadows do traverse the entire outer disc but once a precession the inner and outer disc have the same orientation (briefly resulting in no shadow).

Recent VLT/SPHERE observations of J1604 by \citet{Pinilla:2018gb} with epochs covering multiple years have shown the evolution of narrow lane shadows from a strongly misaligned inner disc. These shadows were found to be confined within $\pm14\degr$ and their direction of travel to apparently reverse across the epochs \citep[e.g. Figure~6 of][]{Pinilla:2018gb}.

Our simulation has demonstrated that a precessing disc can naturally explain how the shadows cast on the outer disc can reverse direction. Additionally, despite not setting out to match any of the disc parameters of J1604 \citep[e.g.][]{Pinilla:2018gb}, the properties of the shadows we find in our simulation are consistent with the observed shadows. Although we found a larger range over which the shadows may be cast (about $\sim 30 \degr$ from the average position, Figure~\ref{fig:disc_with_time}), this characteristic is most strongly influenced by the relative misalignment between the inner and outer disc. This is measured in J1604 to be $\sim 70-90 \degr$ while in our disc this varies between the lower range of $20-70 \degr$. Also, over the course of our simulation the inner disc slowly aligns with the binary, decreasing the range over which the shadow rocks (lower panel, Figure~\ref{fig:disc_with_time}) so we anticipate that the shadows are confined to a smaller range on longer time-scales.

The consistency between the shadow properties in our simulation and those observed in J1604 might suggest that the inner disc is precessing due to a misaligned binary. However, the timescale for the variability found in our simulation is much longer than the day-to-day variability identified in \citet{Pinilla:2018gb}. According to Equation 4 from \citetalias{Facchini:2017of}, the precession timescale of the inner disc is shorter for smaller binary separations or when the inner disc has a smaller radial extent. Recent observations by \citet{Sicilia-Aguilar:2019xx} suggest that the inner disc is more compact than in our simulations. With a smaller inner disc, the precession would be more rapid and the variability of the shadows we find will occur on shorter timescales. 

The recent observations by \citet{Sicilia-Aguilar:2019xx} also find no suggestion of a companion in the inner regions. Previous observations by \citet{Mayama:2018yd} present CO moment maps of J1604, showing a twisted pattern (characteristic of a misaligned inner disc) out to large radii. Previous studies have shown that these twisted kinematics can be readily explained by the presence of a misaligned companion and are consistent with other observational features \citep{Price:2018pf,Calcino:2019bo}.
A misaligned stellar magnetic field, as proposed by \citet{Sicilia-Aguilar:2019xx}, is unlikely to be able to misalign gas on the large scales implied by \citet{Mayama:2018yd}.

As the measured precession timescale of the inner disc depends on both the properties of the binary (separation, relative masses) and the inner disc (extent, outer radius, relative inclination, viscosity), it is not straightforward to infer properties of either from the observed variability. However, the extent of the inner disc can be used to place a limit on the binary separation and mass ratio \citep{Artymowicz:1994uq,nixon_2013} and the shadow cast on the outer disc may be used to estimate the scale-height of the inner disc \citep{Min:2017oc}. Evolution in the CO kinematics may offer additional insights when paired with the scattered light (as in our Figure~\ref{fig:density_sl}).


\section{Conclusion}
We model the evolution of shadows in a broken circumbinary disc. As the inner disc precesses about a direction which is strongly misaligned to the outer disc, the shadows it casts on the outer disc rock and back forth rather than traversing the entire disc. Hence shadows observed in broken discs do not necessarily travel in the same sense as the inner disc precession. These `rocking' shadows have the same period as the inner disc precession and naturally lead to variations in the amplitude and width of the shadows cast. The properties of the shadows cast by the misaligned inner disc are consistent with recent observations of the strongly misaligned disc J1604 \citep{Pinilla:2018gb} but with variation on longer timescales.


\section*{Acknowledgements}
RN thanks Nicol\'{a}s Cuello, Claire Davies and Enrico Ragusa for discussions and Monash University for their hospitality during the visit which initiated this work. This project received funding from the European Research Council (ERC) under the European Union's Horizon 2020 research and innovation programme (grant agreement No 681601) and under the Marie Sk\l{}odowska-Curie grant 823823 (DUSTBUSTERS). We used the DiRAC Data Intensive service at Leicester, operated by the University of Leicester IT Services, part of the STFC DiRAC HPC Facility (www.dirac.ac.uk). Equipment was funded by BEIS capital funding via STFC capital grants ST/K000373/1 and ST/R002363/1 and STFC DiRAC Operations grant ST/R001014/1. DiRAC is part of the National e-Infrastructure. Figures were made using matplotlib \citep{matplotlib}, \textsc{plonk} \citep{Mentiplay:2019pw} and pymcfost.



\bibliographystyle{mnras}
\bibliography{bec}

\begin{thebibliography}{}
\makeatletter
\relax
\def\mn@urlcharsother{\let\do\@makeother \do\$\do\&\do\#\do\^\do\_\do\%\do\~}
\def\mn@doi{\begingroup\mn@urlcharsother \@ifnextchar [ {\mn@doi@}
  {\mn@doi@[]}}
\def\mn@doi@[#1]#2{\def\@tempa{#1}\ifx\@tempa\@empty \href
  {http://dx.doi.org/#2} {doi:#2}\else \href {http://dx.doi.org/#2} {#1}\fi
  \endgroup}
\def\mn@eprint#1#2{\mn@eprint@#1:#2::\@nil}
\def\mn@eprint@arXiv#1{\href {http://arxiv.org/abs/#1} {{\tt arXiv:#1}}}
\def\mn@eprint@dblp#1{\href {http://dblp.uni-trier.de/rec/bibtex/#1.xml}
  {dblp:#1}}
\def\mn@eprint@#1:#2:#3:#4\@nil{\def\@tempa {#1}\def\@tempb {#2}\def\@tempc
  {#3}\ifx \@tempc \@empty \let \@tempc \@tempb \let \@tempb \@tempa \fi \ifx
  \@tempb \@empty \def\@tempb {arXiv}\fi \@ifundefined
  {mn@eprint@\@tempb}{\@tempb:\@tempc}{\expandafter \expandafter \csname
  mn@eprint@\@tempb\endcsname \expandafter{\@tempc}}}

\bibitem[\protect\citeauthoryear{{Artymowicz} \& {Lubow}}{{Artymowicz} \&
  {Lubow}}{1994}]{Artymowicz:1994uq}
{Artymowicz} P.,  {Lubow} S.~H.,  1994, \mn@doi [\apj] {10.1086/173679}, \href
  {http://adsabs.harvard.edu/abs/1994ApJ...421..651A} {421, 651}

\bibitem[\protect\citeauthoryear{{Bate}, {Bonnell}, {Clarke}, {Lubow},
  {Ogilvie}, {Pringle}  \& {Tout}}{{Bate} et~al.}{2000}]{Bate:2000fk}
{Bate} M.~R.,  {Bonnell} I.~A.,  {Clarke} C.~J.,  {Lubow} S.~H.,  {Ogilvie}
  G.~I.,  {Pringle} J.~E.,   {Tout} C.~A.,  2000, \mn@doi [\mnras]
  {10.1046/j.1365-8711.2000.03648.x}, \href
  {http://adsabs.harvard.edu/abs/2000MNRAS.317..773B} {317, 773}

\bibitem[\protect\citeauthoryear{{Benisty} et~al.,}{{Benisty}
  et~al.}{2018}]{Benisty:2018ve}
{Benisty} M.,  et~al., 2018, \mn@doi [\aap] {10.1051/0004-6361/201833913},
  \href {https://ui.adsabs.harvard.edu/abs/2018A&A...619A.171B} {619, A171}

\bibitem[\protect\citeauthoryear{{Biller} et~al.,}{{Biller}
  et~al.}{2012}]{Biller:2012ct}
{Biller} B.,  et~al., 2012, \mn@doi [\apjl] {10.1088/2041-8205/753/2/L38},
  \href {https://ui.adsabs.harvard.edu/abs/2012ApJ...753L..38B} {753, L38}

\bibitem[\protect\citeauthoryear{{Calcino}, {Price}, {Pinte}, {van der Marel},
  {Ragusa}, {Dipierro}, {Cuello}  \& {Christiaens}}{{Calcino}
  et~al.}{2019}]{Calcino:2019bo}
{Calcino} J.,  {Price} D.~J.,  {Pinte} C.,  {van der Marel} N.,  {Ragusa} E.,
  {Dipierro} G.,  {Cuello} N.,   {Christiaens} V.,  2019, \mn@doi [\mnras]
  {10.1093/mnras/stz2770}, \href
  {https://ui.adsabs.harvard.edu/abs/2019MNRAS.490.2579C} {490, 2579}

\bibitem[\protect\citeauthoryear{{Casassus} et~al.,}{{Casassus}
  et~al.}{2015}]{Casassus:2015yu}
{Casassus} S.,  et~al., 2015, \mn@doi [\apj] {10.1088/0004-637X/811/2/92},
  \href {http://adsabs.harvard.edu/abs/2015ApJ...811...92C} {811, 92}

\bibitem[\protect\citeauthoryear{Casassus et~al.,}{Casassus
  et~al.}{2018}]{Casassus:2018te}
Casassus S.,  et~al., 2018, \mn@doi [\mnras] {10.1093/mnras/sty894}, 477, 5104

\bibitem[\protect\citeauthoryear{{Casassus}, {P{\'e}rez}, {Osses}  \&
  {Marino}}{{Casassus} et~al.}{2019}]{Casassus:2019pt}
{Casassus} S.,  {P{\'e}rez} S.,  {Osses} A.,   {Marino} S.,  2019, \mn@doi
  [\mnras] {10.1093/mnrasl/slz059}, \href
  {https://ui.adsabs.harvard.edu/abs/2019MNRAS.486L..58C} {486, L58}

\bibitem[\protect\citeauthoryear{{Facchini}, {Juh{\'a}sz}  \&
  {Lodato}}{{Facchini} et~al.}{2018}]{Facchini:2017of}
{Facchini} S.,  {Juh{\'a}sz} A.,   {Lodato} G.,  2018, \mn@doi [\mnras]
  {10.1093/mnras/stx2523}, \href
  {https://ui.adsabs.harvard.edu/\#abs/2018MNRAS.473.4459F} {473, 4459}

\bibitem[\protect\citeauthoryear{{Flebbe}, {Muenzel}, {Herold}, {Riffert}  \&
  {Ruder}}{{Flebbe} et~al.}{1994}]{Flebbe:1994lr}
{Flebbe} O.,  {Muenzel} S.,  {Herold} H.,  {Riffert} H.,   {Ruder} H.,  1994,
  \mn@doi [\apj] {10.1086/174526}, \href
  {http://adsabs.harvard.edu/abs/1994ApJ...431..754F} {431, 754}

\bibitem[\protect\citeauthoryear{Hunter}{Hunter}{2007}]{matplotlib}
Hunter J.~D.,  2007, \mn@doi [Computing In Science \& Engineering]
  {10.1109/MCSE.2007.55}, 9, 90

\bibitem[\protect\citeauthoryear{{Lacour} et~al.,}{{Lacour}
  et~al.}{2016}]{Lacour:2016ij}
{Lacour} S.,  et~al., 2016, \mn@doi [\aap] {10.1051/0004-6361/201527863}, \href
  {https://ui.adsabs.harvard.edu/abs/2016A&A...590A..90L} {590, A90}

\bibitem[\protect\citeauthoryear{{Marino}, {Perez}  \& {Casassus}}{{Marino}
  et~al.}{2015}]{Marino:2015rh}
{Marino} S.,  {Perez} S.,   {Casassus} S.,  2015, \mn@doi [\apjl]
  {10.1088/2041-8205/798/2/L44}, \href
  {http://adsabs.harvard.edu/abs/2015ApJ...798L..44M} {798, L44}

\bibitem[\protect\citeauthoryear{{Mathis}, {Rumpl}  \& {Nordsieck}}{{Mathis}
  et~al.}{1977}]{Mathis:1977bw}
{Mathis} J.~S.,  {Rumpl} W.,   {Nordsieck} K.~H.,  1977, \mn@doi [\apj]
  {10.1086/155591}, \href
  {https://ui.adsabs.harvard.edu/#abs/1977ApJ...217..425M} {217, 425}

\bibitem[\protect\citeauthoryear{{Mayama} et~al.,}{{Mayama}
  et~al.}{2018}]{Mayama:2018yd}
{Mayama} S.,  et~al., 2018, \mn@doi [\apjl] {10.3847/2041-8213/aae88b}, \href
  {https://ui.adsabs.harvard.edu/abs/2018ApJ...868L...3M} {868, L3}

\bibitem[\protect\citeauthoryear{{Mentiplay}}{{Mentiplay}}{2019}]{Mentiplay:2019pw}
{Mentiplay} D.,  2019, \mn@doi [The Journal of Open Source Software]
  {10.21105/joss.01884}, \href
  {https://ui.adsabs.harvard.edu/abs/2019JOSS....4.1884M} {4, 1884}

\bibitem[\protect\citeauthoryear{{Min}, {Stolker}, {Dominik}  \&
  {Benisty}}{{Min} et~al.}{2017}]{Min:2017oc}
{Min} M.,  {Stolker} T.,  {Dominik} C.,   {Benisty} M.,  2017, \mn@doi [\aap]
  {10.1051/0004-6361/201730949}, \href
  {http://ukads.nottingham.ac.uk/abs/2017A%26A...604L..10M} {604, L10}

\bibitem[\protect\citeauthoryear{{Nealon}, {Dipierro}, {Alexander}, {Martin}
  \& {Nixon}}{{Nealon} et~al.}{2018}]{Nealon:2018ic}
{Nealon} R.,  {Dipierro} G.,  {Alexander} R.,  {Martin} R.~G.,   {Nixon} C.,
  2018, \mn@doi [\mnras] {10.1093/mnras/sty2267}, \href
  {https://ui.adsabs.harvard.edu/\#abs/2018MNRAS.481...20N} {481, 20}

\bibitem[\protect\citeauthoryear{{Nixon}, {King}  \& {Price}}{{Nixon}
  et~al.}{2013}]{nixon_2013}
{Nixon} C.,  {King} A.,   {Price} D.,  2013, \mn@doi [\mnras]
  {10.1093/mnras/stt1136}, \href
  {http://adsabs.harvard.edu/abs/2013MNRAS.434.1946N} {434, 1946}

\bibitem[\protect\citeauthoryear{{Pinilla} et~al.,}{{Pinilla}
  et~al.}{2018}]{Pinilla:2018gb}
{Pinilla} P.,  et~al., 2018, \mn@doi [\apj] {10.3847/1538-4357/aae824}, \href
  {https://ui.adsabs.harvard.edu/abs/2018ApJ...868...85P} {868, 85}

\bibitem[\protect\citeauthoryear{{Pinte}, {M{\'e}nard}, {Duch{\^e}ne}  \&
  {Bastien}}{{Pinte} et~al.}{2006}]{Pinte:2006nw}
{Pinte} C.,  {M{\'e}nard} F.,  {Duch{\^e}ne} G.,   {Bastien} P.,  2006, \mn@doi
  [\aap] {10.1051/0004-6361:20053275}, \href
  {http://adsabs.harvard.edu/abs/2006A%26A...459..797P} {459, 797}

\bibitem[\protect\citeauthoryear{{Pinte}, {Harries}, {Min}, {Watson},
  {Dullemond}, {Woitke}, {M{\'e}nard}  \& {Dur{\'a}n-Rojas}}{{Pinte}
  et~al.}{2009}]{Pinte:2009ye}
{Pinte} C.,  {Harries} T.~J.,  {Min} M.,  {Watson} A.~M.,  {Dullemond} C.~P.,
  {Woitke} P.,  {M{\'e}nard} F.,   {Dur{\'a}n-Rojas} M.~C.,  2009, \mn@doi
  [\aap] {10.1051/0004-6361/200811555}, \href
  {http://adsabs.harvard.edu/abs/2009A%26A...498..967P} {498, 967}

\bibitem[\protect\citeauthoryear{{Price} et~al.,}{{Price}
  et~al.}{2018a}]{Phantom}
{Price} D.~J.,  et~al., 2018a, \mn@doi [\pasa] {10.1017/pasa.2018.25}, \href
  {https://ui.adsabs.harvard.edu/\#abs/2018PASA...35...31P} {35, e031}

\bibitem[\protect\citeauthoryear{{Price} et~al.,}{{Price}
  et~al.}{2018b}]{Price:2018pf}
{Price} D.~J.,  et~al., 2018b, \mn@doi [\mnras] {10.1093/mnras/sty647}, \href
  {http://ukads.nottingham.ac.uk/abs/2018MNRAS.477.1270P} {477, 1270}

\bibitem[\protect\citeauthoryear{{Sicilia-Aguilar}, {Manara}, {de Boer},
  {Benisty}, {Pinilla}  \& {Bouvier}}{{Sicilia-Aguilar}
  et~al.}{2019}]{Sicilia-Aguilar:2019xx}
{Sicilia-Aguilar} A.,  {Manara} C.~F.,  {de Boer} J.,  {Benisty} M.,  {Pinilla}
  P.,   {Bouvier} J.,  2019, arXiv e-prints, \href
  {https://ui.adsabs.harvard.edu/abs/2019arXiv191104938S} {p. arXiv:1911.04938}

\bibitem[\protect\citeauthoryear{{Siess}, {Dufour}  \& {Forestini}}{{Siess}
  et~al.}{2000}]{Siess:2000vd}
{Siess} L.,  {Dufour} E.,   {Forestini} M.,  2000, \aap, \href
  {https://ui.adsabs.harvard.edu/#abs/2000A&A...358..593S} {358, 593}

\bibitem[\protect\citeauthoryear{{Walsh}, {Daley}, {Facchini}  \&
  {Juh{\'a}sz}}{{Walsh} et~al.}{2017}]{Walsh:2017ic}
{Walsh} C.,  {Daley} C.,  {Facchini} S.,   {Juh{\'a}sz} A.,  2017, \mn@doi
  [\aap] {10.1051/0004-6361/201731334}, \href
  {https://ui.adsabs.harvard.edu/\#abs/2017A&A...607A.114W} {607, A114}

\bibitem[\protect\citeauthoryear{{Zhu}}{{Zhu}}{2019}]{Zhu:2018vf}
{Zhu} Z.,  2019, \mn@doi [\mnras] {10.1093/mnras/sty3358}, \href
  {https://ui.adsabs.harvard.edu/abs/2019MNRAS.483.4221Z} {483, 4221}

\bibitem[\protect\citeauthoryear{{van der Plas} et~al.,}{{van der Plas}
  et~al.}{2019}]{vanderPlas:2019gy}
{van der Plas} G.,  et~al., 2019, \mn@doi [\aap] {10.1051/0004-6361/201834134},
  \href {https://ui.adsabs.harvard.edu/abs/2019A&A...624A..33V} {624, A33}

\makeatother
\end{thebibliography}

\bsp	
\label{lastpage}
\end{document}